\documentclass[a4,useAMS,usenatbib,usegraphicx]{mn2e}
\usepackage{subfig}
\usepackage{lscape}
\usepackage{natbib}
\usepackage{amssymb,amsmath}
\usepackage{fixltx2e}
\title[The Arecibo Galaxy Environment Survey VI : The Virgo Cluster (II)]{The Arecibo Galaxy Environment Survey VI : The Virgo Cluster (II)}
\author[R. Taylor, J. I. Davies, R. Auld, R. F. Minchin, R. Smith]{R. Taylor$^{1}$\thanks{Email: rhyst@naic.edu}, J. I. Davies$^2$, R. Auld$^2$, R. F. Minchin$^1$, R. Smith$^{3}$.\\
$^1$Arecibo Observatory, HC03 Box 53995, Arecibo, Puerto Rico 00612\\
$^2$School of Physics \& Astronomy, Cardiff University, Queens Buildings, The Parade, Cardiff CF24 3AA, U.K.\\
$^3$Department of Astronomy, Esteban Iturra Avenue, Universidad de Concepci\'{o}n, Casilla (P.O.BOX) 160-C, Concepci\'{o}n, Chile.}
\begin{document}

\date{Submitted July 2011}

\pagerange{\pageref{firstpage}--\pageref{lastpage}} \pubyear{2011}

\maketitle

\label{firstpage}

\begin{abstract}
We present 21 cm observations of a 5 $\times$ 1 degree region in the Virgo cluster, obtained as part of the Arecibo Galaxy Environment Survey. 13 cluster members are detected, together with 36 objects in the background. We compare and contrast the results from this area with a larger 10 $\times$ 2 degree region. We combine the two data sets to produce an HI mass function, which shows a higher detection rate at low masses (but finds fewer massive galaxies) than less sensitive wider-area surveys, such as ALFALFA. We find that the HI-detected galaxies are distributed differently to the non-detections, both spatially and in velocity, providing further evidence that the cluster is still assembling. We use the Tully-Fisher relation to examine the possibility of morphological evolution. We find that highly deficient galaxies, as well as some early-type galaxies, have much lower velocity widths than the Tully-Fisher relation predicts, indicating gas loss via ram pressure stripping. We also find that HI detections without optical counterparts do not fit the predictions of the baryonic Tully-Fisher relation, implying that they are not primordial objects.
\end{abstract}


\section{Introduction}

\begin{figure}
\includegraphics[width=84mm]{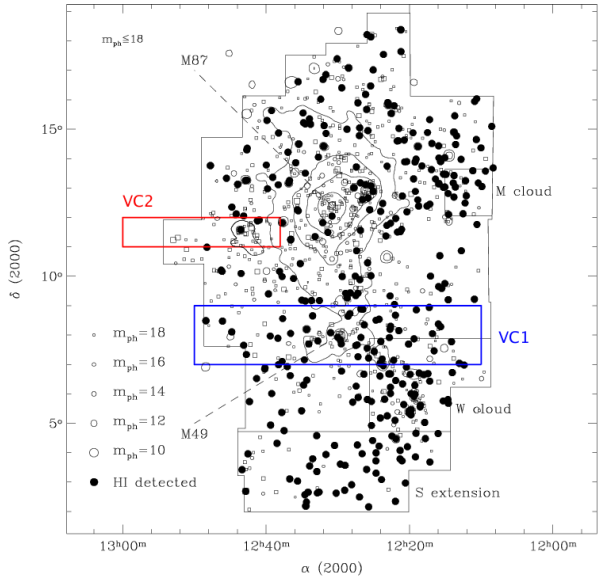}
\caption[Virgo Cluster map showing the AGES fields]{The Virgo Cluster region, highlighting the AGES and VCC areas and detections. Squares are early-type while circles are late-type galaxies. Filled circles are those with known HI detections from previous pointed observations. The contours show X-ray detection from the ROSAT satellite. Physically the VC1 area spans 3.0$\times$0.6 Mpc while VC2 spans 1.5$\times$0.3 Mpc (at 17 Mpc distance). This figure is reproduced from paper V.}
\label{VMap}
\end{figure}

We previously described the results of a 10 $\times$ 2 degree survey of the Virgo Cluster with AGES, the Arecibo Galaxy Environment Survey (\citealt{me}, hereafter Paper V). This fully-sampled HI survey reaches a mass sensitivity of $\sim$8$\times$10$^{6}$ $M_{\rm \odot}$ at the Virgo distance of 17 Mpc. However, the cluster is a complex and varied environment. While our first region (VC1) covered the dense cluster interior around M49, where multiple sub-clusters are infalling, our second, smaller area (VC2 measures 5 $\times$ 1 degree), examines a less dense area on the cluster periphery (see figure \ref{VMap}). In this more quiescent environment, galaxies are thought to belong to a  population at a single distance (\citealt{gv99}). Our two areas are thus very different, so understanding both is important in understanding the cluster as a whole. 

We described in Paper V evidence that the Virgo Cluster is still in the process of assembly. HI-detected galaxies are found preferentially at higher velocities than galaxies devoid of HI, implying that infalling galaxies have not yet experienced the full force of the cluster environment. We also found that virtually all late-type galaxies (listed in the Virgo Cluster Catalogue of \citealt{bst}) are detected in HI, even if they are strongly HI deficient, suggesting that gas loss can cause morphological evolution in the nearby Universe. 

Further evidence of this is suggested by the relative dearth of early-type detections. No more than 9\% of early-type galaxies are detected in HI, but if this is restricted to S0s (which are bluer than other early-type galaxies and show some evidence of structure) the fraction rises to 25\%. Conversely, the fraction of detected dEs (which are redder and do not show evidence of structure) is only 3\%. Despite employing a variety of stacking procedures (which reached a mass sensitivity about 6 times better than the standard survey data), the majority of early-types remain undetected in HI.

At the other extreme, 8 detections in the VC1 area (about 7\% of the sample) have no obvious optical counterparts. Such objects may represent either tidal debris, very low surface brightness galaxies, or, more controversially, optically dark galaxies. In either case, most of the HI detections not associated with previously known VCC galaxies correspond to faint, blue non-deficient dwarf galaxies. Clearly there is at least some prospect for the survival of low mass, gas-rich and optically faint objects within the cluster environment. 

In this paper, we add to our analysis of the detections in VC1 and also describe the detections in VC2. We seek to further our understanding of the issues outlined above in two ways. Firstly, studying a second area increases our sample size for better overall statistics. Secondly, existing observations indicate that this area should be quite different to the first, with galaxies belonging to a population at a single line-of-sight distance (as opposed to clouds at 3 different distances in VC1) and less densely distributed. In VC1 there are about 17 VCC galaxies per square degree, but only 13 per square degree in VC2. We therefore hope to evaluate our conclusions from Paper V in this quite different environment.

Galaxies within the VC2 region are all thought to be part of subcluster A and are assigned a distance of 17 Mpc in the Galaxy On-Line Milano Network database (GOLDMine, \citealt{gv03}). The central giant elliptical galaxy M87 is avoided, as its strong continuum emission can cause severe interference by setting up strong standing waves. The field extends beyond the edge of the VCC plates (see figure \ref{VMap}), so it is hoped that changes in galaxy properties from the cluster interior to the field may be seen. Physically the area spans 1.5$\times$0.3 Mpc at 17 Mpc distance.

The remainder of this paper is as follows. In section \ref{sec:Obs} we describe the observations, data reduction and analysis procedures. In section \ref{sec:VCOs} we describe our results for cluster members.  In section \ref{sec:Conclusions} we summarise and describe some of the implications of this survey within the Virgo cluster, while we comment on some of our background detections in the Appendix.

\section{Data analysis overview}
\label{sec:Obs}

The observations, data reduction and analysis procedures have been fully described in \citealt{a06}, as well as in \citealt{c08}, \citealt{m10} and in Paper V. For this reason we here give only a very brief summary. The field consists of an area of uniform sensitivity 5 degrees in R.A. by 1 degree in declination; due to the hexagonal beam pattern of ALFA a small area outside this field is also included. The maximum range of the field in R.A. is from 12:37:35 to 12:58:52; in declination it ranges from +10:56:00 to +12:13:00. The maximum velocity range is from -2,220 to 20,090 km/s (including the bandpass roll-off where sensitivity is reduced).

The HI data are reduced with the \textsc{aips++} packages \textsc{livedata} and \textsc{gridzilla}, and analysed using \textsc{miriad}, primarily through the task \textsc{mbspect}. ALFA records data from two linear polarisations simultaneously and we reduce the data for both of these separately, as the noise between each is uncorrelated (see next section). We also reduce the combination of both polarisations - it is on this cube that we perform our actual measurements. For full details of data reduction and our analysis techniques, we refer the reader to Paper V.

\subsection{Source extraction}

Our source extraction procedures are largely identical to those described in Paper V - namely, repeated visual searches of the cube coupled with our own custom-made automatic extractor (\textsc{glados}) described in that paper. In this region we also employed the existing extractors \textsc{polyfind} (see \citealt{poly}) and \textsc{duchamp} (described in \citealt{duch}). Here we present a brief comparison of the relative performance and efficiency of each technique.

\textsc{polyfind}, which was designed for HIPASS data, proved very unreliable - generating a source catalogue of over 1,600 candidates, of which only about 3\% proved to be real (the majority dismissed as single-channel, relatively low S/N spikes visible in only one polarisation). \textsc{duchamp} fares rather better with AGES data, but as a flux (rather than S/N) based extractor, it has major problems at the edges of AGES cubes (where the noise is significantly stronger). Unfortunately in the particular case of VC2, a number of very bright sources are present near the edges, which \textsc{duchamp} has to either reject or accept an even lower level of reliability than \textsc{polyfind}.

\textsc{glados}, which is more fully described in Paper V, is an automatic extractor designed to reproduce selected features of both \textsc{polyfind} and \textsc{duchamp}. While it checks for peak S/N values along individual spectra, as \textsc{polyfind} does, it also checks that those peaks span multiple pixels, in the manner of \textsc{duchamp}. Additionally, it is designed to check both of the polarisations separately, only accepting a source if it meets the detection criteria in all 3 data cubes. For the other extractors we check individually each detection candidate to see if they are visible in both polarisations.

Table \ref{SEPT} presents completeness and reliability estimates for \textsc{polyfind}, \textsc{duchamp}, \textsc{glados} and a visual search. We can never know exactly how many HI sources are really present within the volume of the data cube, so the completeness is estimated by considering the source catalogue resulting from a combination of all search techniques. Thus a completeness estimate of 50\% means that a particular search technique was able to find 50\% of all detectable sources. A major additional caveat is that the sensitivity levels are not identical for each method - this table is only intended as an approximate comparison.

\begin{table}
\footnotesize
\begin{center}
\caption[Percentage reliability and completeness estimates for source extraction methods]{Percentage reliability and completeness estimates for various source extraction methods. The completeness value in brackets is the completeness based on the volume of the cube each method is able to search - see text for details.}
\label{SEPT}
\begin{tabular}{c c c c}\\
\hline
  \multicolumn{1}{c}{Method} &
  \multicolumn{1}{c}{No. candidates} &
  \multicolumn{1}{c}{Completeness} &
  \multicolumn{1}{c}{Reliability} \\
\hline
  Visual & 75 & 77(77) & 51\\
  \textsc{polyfind} & 1619 & 70(73) & 2\\
  \textsc{duchamp} & 492 & 72(90) & 7\\
  \textsc{glados} & 268 & 83(95) & 16\\
\hline
\end{tabular}
\end{center}
\end{table}

In table \ref{SEPT} we provide two estimates of the completeness. The first is simply the percentage of the final source list found by each method. The second accounts for the fact that automatic extractors are not able to easily search the entire volume of the cube. All automatic extractors cope badly in regions affected by RFI. These regions are therefore generally masked in order to avoid generating candidate lists numbering in the thousands, which are impractical to search. 

We note that \textsc{duchamp} also has a further constraint - to determine a sensible flux threshold as a search parameter, it is necessary to have already searched the cube in some way. If we instead use \textsc{duchamp's} ability to estimate the S/N level, the reliability is raised to 18\%, but with a huge decrease in completeness to just 42\%, or 54\% in the volume it is able to search.

The reliability estimate is made by considering what fraction of the detected sources are real. As in Paper V, a source is accepted as a candidate if it is detected in both polarisations and by multiple methods of source extraction. Sources are subject to follow-up observations with L-wide if they are only detected by 1 method (but appear in both polarisations), have no clear optical counterpart, or are otherwise anomalous in some way (such as having very different parameters to those given in the literature).

\subsection{Other data}

Our procedures for optical data analysis are very similar to those described in Paper V. We primarily use the Sloan Digital Sky Survey (SDSS) to search for optical counterparts within a 3.5$\arcmin$ radius of the HI coordinates. Objects which are within the Arecibo beam but do not have optical redshift data are flagged as uncertain counterparts. We treat these as being associated with the HI throughout the analysis, except where there are multiple candidates within the beam. 

For this region there is also Isaac Newton Telescope Wide Field Survey (INT WFS) data available. Unlike VC1, in this case the data is uniformly of high quality and slightly deeper than the SDSS (by about 1 mag arcsec$^{-2}$). However, the INT $i$ band images are subject to varying levels of ``fringing'' which is greatly reduced  - but not entirely removed - by the automatic pipeline (\citealt{pipe}). This can be a severe problem for faint, low surface brightness sources. For this reason, and also to maintain consistency with paper V, our photometry is again performed using the SDSS data.

As in Paper V we checked how closely the positions, recessional velocities and velocity width estimates for our data agree with our sources (NED, the SDSS and GOLDMine). We find the agreements to be excellent - the mean difference in the positions of the HI and optical coordinates is  23", the mean difference in recessional velocities is 4 km/s, and the difference in velocity widths is 10 km/s.

\section{Virgo Cluster Objects}
\label{sec:VCOs}

Table \ref{HIVirgo} lists the HI parameters of the objects identified as Virgo cluster members\footnote{This table, together with the HI data for all HI detections behind the cluster, will be available electronically through the Virtual Observatory.}. All quantities are measured/calculated as described in Paper V. Table \ref{SDSSVirgo} lists all of our measured optical parameters for Virgo objects; those not included have multiple or unclear optical counterparts.

\begin{table*}
\tiny
\begin{center}
\caption[HI properties of AGES detections within the Virgo Cluster in VC2]{HI properties of Virgo Cluster galaxies. Bracketed values indicate errors as estimated by \textsc{miriad}. Columns : (1) Source number in this catalogue (2) Right ascension J2000, error in seconds of time (3) Declination J2000, error in arcseconds (4) Heliocentric velocity km/s (5) Maximum velocity width at 50$\%$ and (6) 20$\%$ of the peak flux (7) Total flux Jy (8) Estimated HI mass, log  $M_{\rm \odot}$ (9) Peak signal to noise (10) R.m.s. mJy.}
\label{HIVirgo}
\begin{tabular}{c c c c c c c c c c}\\
\hline
  \multicolumn{1}{c}{(1) No.} &
  \multicolumn{1}{c}{(2) R.A.} &
  \multicolumn{1}{c}{(3) Dec.} &
  \multicolumn{1}{c}{(4) Velocity (km/s)} &
  \multicolumn{1}{c}{(5) W50 (km/s)} &
  \multicolumn{1}{c}{(6) W20 (km/s)} &
  \multicolumn{1}{c}{(7) Flux (Jy)} &
  \multicolumn{1}{c}{(8) M$_{HI}$} &
  \multicolumn{1}{c}{(9) S/N} &
  \multicolumn{1}{c}{(10) rms} \\
\hline
  AGESVC2\_18 & 12:41:12.0(0.7) & +11:53:13(10) & 2259(2) & 227(4) & 248(6) & 3.614(0.190) & 8.34 & 26.69 & 0.6\\
  AGESVC2\_19 & 12:43:07.4(0.7) & +12:03:21(12) & 2026(3) & 41(6) & 84(10) & 0.537(0.078) & 7.51 & 19.31 & 0.7\\
  AGESVC2\_20 & 12:48:11.0(1.3) & +10:58:19(15) & 1163(4) & 78(8) & 116(12) & 3.785(0.531) & 8.36 & 12.62 & 4.2\\
  AGESVC2\_22 & 12:43:31.6(0.7) & +11:34:51(10) & 1418(2) & 173(3) & 195(5) & 6.067(0.317) & 8.56 & 62.70 & 0.6\\
  AGESVC2\_23 & 12:37:44.0(0.9) & +11:49:20(13) & 1537(3) & 331(6) & 368(9) & 7.906(0.551) & 8.68 & 19.36 & 2.4\\
  AGESVC2\_24 & 12:44:58.1(0.7) & +12:01:56(10) & 1544(2) & 37(3) & 52(5) & 0.538(0.067) & 7.51 & 27.53 & 0.5\\
  AGESVC2\_25 & 12:51:05.9(0.7) & +12:03:33(10) & 1793(1) & 39(3) & 54(4) & 1.466(0.157) & 7.95 & 63.80 & 0.6\\
  AGESVC2\_27 & 12:51:55.0(0.7) & +12:05:00(10) & 1785(2) & 334(3) & 358(5) & 12.151(0.476) & 8.87 & 59.90 & 0.7\\
  AGESVC2\_29 & 12:44:08.9(0.7) & +12:07:05(10) & 1013(1) & 106(3) & 122(4) & 4.039(0.279) & 8.39 & 44.53 & 0.9\\
  AGESVC2\_30 & 12:38:39.8(0.7) & +11:58:46(10) & 1041(3) & 42(6) & 74(9) & 0.402(0.063) & 7.39 & 16.67 & 0.6\\ 
  AGESVC2\_33 & 12:40:55.2(0.8) & +11:55:03(11) & 1653(6) & 115(12) & 195(18) & 0.666(0.085) & 7.60 & 11.80 & 0.6\\
  AGESVC2\_35 & 12:56:43.2(0.7) & +11:55:52(10) & 569(1) & 57(3) & 76(4) & 1.847(0.169) & 8.05 & 92.58 & 0.4\\
  AGESVC2\_63 & 12:56:04.9(0.8) & +12:09:07(17) & 1869(10) & 32(20) & 89(30) & 0.219(0.092) & 7.55 & 5.25 & 1.2\\
\hline
\end{tabular}
\end{center}
\end{table*}

\begin{table*}
\tiny
\begin{center}
\caption{Optical properties of HI detections in the Virgo Cluster. Columns : (1) Source number in this catalogue, (2) Flag for status of the optical counterpart used here - 0 indicates the object has a matching optical redshift, 1 indicates no optical redshift is available (3) Name in a major catalogue if present (4) Morphological type, using the system of the GOLDMine database (5) Absolute magnitude in the $g$ and (6) $i$ bands (7) $g$-$i$ colour (8) HI mass-to-light ratio, $g$ band (9) Difference between optical and HI centres in arcminutes (10) Optical diameter in kpc (11) Estimated HI deficiency - see equation \ref{DefEqt}. An * indicates the INT $B$-band was used for photometry.}
\label{SDSSVirgo}
\begin{tabular}{c c c c c c c c c c c}\\
\hline
  \multicolumn{1}{c}{(1) No.} &
  \multicolumn{1}{c}{(2) Flag} &
  \multicolumn{1}{c}{(3) Name} &
  \multicolumn{1}{c}{(4) MType} &
  \multicolumn{1}{c}{(5) M$_{g}$} &
  \multicolumn{1}{c}{(6) M$_{i}$} &
  \multicolumn{1}{c}{(7) g-i} &
  \multicolumn{1}{c}{(8) $M_{\rm HI}$/$L_{\rm g}$} &
  \multicolumn{1}{c}{(9) Separation} &
  \multicolumn{1}{c}{(10) OptD} &
  \multicolumn{1}{c}{(11) $HI_{def}$} \\
\hline
  AGESVC2\_18 & 0 & VCC1868 & 8(Scd) & -17.82 & -18.99 & 1.17 & 0.14 & 0.04 & 14.91 & 0.70\\
  AGESVC2\_19 & 0 & VCC1955 & 14(S/BCD) & -17.68 & -18.58 & 0.89 & 0.02 & 0.29 & 17.03 & 1.61\\
  AGESVC2\_23 & 0 & VCC1727 & 4(Sab) & -21.06 & -22.22 & 1.16 & 0.01 & 0.28 & 19.46 & 0.73\\
  AGESVC2\_24 & 1 & AGC225876& 20(?) & -11.83 & -12.06 & 0.23 & 5.01 & 0.12 & 1.89 & 0.27\\
  AGESVC2\_25 & 0 & AGC223205  & 19(S) & -13.87 & -14.19 & 0.33 & 2.10 & 0.22 & 5.03 & 0.43\\
  AGESVC2\_27 & 0 & NGC4746 & 5(Sb) & -18.13 & -19.06 & 0.93 & 0.34 & 0.04 & 19.46 & 0.50\\
  AGESVC2\_29 & 0 & VCC1992 & 12(Im) & -15.61 & -16.06 & 0.45 & 1.16 & 0.31 & 10.96 & 0.47\\
  AGESVC2\_30 & 0 & None & 20(?) & -10.54* & - & - & 9.86* & 0.23 & 2.84 & 0.80\\
  AGESVC2\_33 & 0 & VCC1859 & 3(Sa) & -18.79 & -19.77 & 0.98 & 0.01 & 0.65 & 28.21 & 1.8\\
  AGESVC2\_35 & 0 & UGC08061 & 12(Im) & -15.22 & -15.83 & 0.60 & 0.76 & 0.12 & 8.58 & 0.66\\
  AGESVC2\_63 & 1 & AGC225878& 20(?) & -10.99 & -11.60 & 0.61 & 4.92 & 1.21 & 1.72 & 0.61\\
\hline
\end{tabular}
\end{center}
\end{table*}

It is immediately apparent that within the cluster, there are far fewer detections in VC2 than in VC1. This is not simply because VC1 is larger : there are about 2.6 HI detections per square degree in VC2 compared with 4.8 per square degree in VC1. It may also be noted from table \ref{HIVirgo} that the S/N estimate is high for all Virgo detections, with only one source with a peak S/N of less than 10 (all the HI detections here are also detected by ALFALFA). This is not a source extraction issue - lower S/N detections in the cluster were originally recorded, but all were rejected based on follow-up observations.

Moreover, the difference in galaxy numbers between VC1 and VC2 is not unusual but simply the varying nature of galaxy density in the cluster - in the 5$\times$1 degree strip immediately west of VC2 there are 23.6 VCC galaxies per square degree. The key point of the lower number of detections is that any kind of detailed comparison between VC1 and VC2 will suffer from small number statistics.

\subsection{The HI mass distribution}

As mentioned above the S/N of the HI detections in VC2 are all relatively high. This is reflected in the high HI masses of the galaxies detected in this region, with only one lower than 3$\times$10$^{7}$  $M_{\rm \odot}$ (7.5 in logarithmic solar units). 23 galaxies below this limit were detected in VC1 which covers an area four times the size of VC2. Given the higher galaxy density in VC1, more low-mass galaxies should be expected there simply because there are more galaxies present overall. If galaxies of low HI-mass were detected in VC2 in the same proportion as in VC1 (as a fraction of the total number of detections in the region), 3 detections would be expected as opposed to the 1 that is actually found - so the number of low-mass detections in the two areas is consistent given $\sqrt{n}$ errors.

\begin{figure}
\begin{center}
\includegraphics[width=80mm]{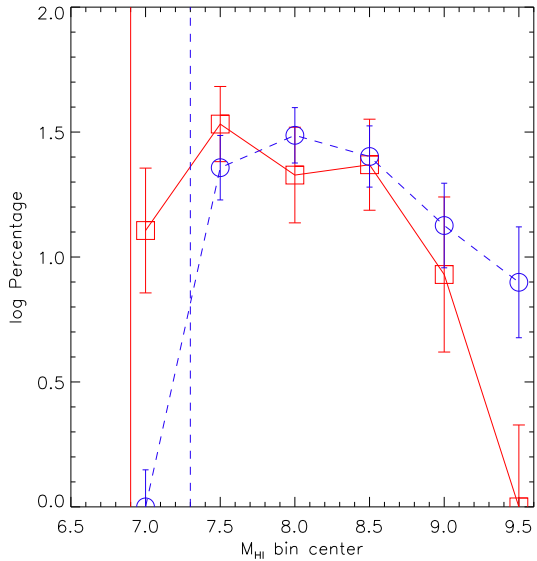}
\caption[HI mass functions from AGES and ALFALFA]{HI mass functions for 47 AGES galaxies at 17 Mpc (solid red line), and the 202 galaxies in ALFALFA within the limits of the VCC (blue dashed line). Vertical lines indicate the sensitivity limits. The x-axis is the HI mass bin center in logarithmic solar units with a bin size of 0.5 (i.e. the 7.0 bin includes galaxies from 6.75 to 7.25). The y-axis is the logarithm of the percentage of the galaxies in each sample in each bin. Errors are 1$\sigma$ (Poisson statistics).}
\label{HIMFs}
\end{center}
\end{figure}

Directly comparing the HI mass distributions between VC1 and VC2 is very difficult since there are only 13 objects in the cluster in VC2, and clouds at 3 different distances (and so differing mass sensitivity limits) within VC1. It is more useful to collate all the HI detections and produce an overall HI mass distribution, using, for consistency with VC2, only those objects at 17 Mpc from VC1. This gives a total sample size of 47 objects. The mass distribution is shown in figure \ref{HIMFs}, which compares the AGES result with that of the ALFALFA survey (\citealt{a40}) in the Virgo region.

Since the surveys cover very different areas and the cluster's overall structure is complex (see Paper V and also figure \ref{VMap}) it is important to ensure a consistent distance estimate for all objects. For both surveys, we reject galaxies located in those areas that \citealt{gv99} have assigned as containing infalling clouds (i.e. galaxies at 23 and 32 Mpc distance). For the remaining galaxies we assume a uniform distance of 17 Mpc, rather than the more complex velocity flow models used by ALFALFA (see \citealt{a40}). Additionally, only `code 1' galaxies from the ALFALFA catalogue are used to ensure that only real objects are included, giving an ALFALFA sample size of 202 objects.

As discussed earlier, the sensitivity limit for AGES noted in figure \ref{HIMFs} assumes a 4$\sigma$, 50 km/s velocity width tophat profile detection. There are some important caveats to this. The survey is noise-limited, not flux limited : the narrower the observed velocity width of a galaxy, the higher the peak S/N for any given total integrated flux. Sources with lower intrinsic rotational velocities, or at lower inclination angles, will thus be easier to detect at low masses. Since visual source extraction (and automatic extractors in the cases of \textsc{glados} and \textsc{polyfind}) relies on S/N, not total integrated flux, the sensitivity limit is only approximate at best. As discussed earlier, estimating the completeness limit is considerably more difficult (see also \citealt{schneider}).

For ALFALFA the sensitivity limit assumes a 4.5$\sigma$ limit as estimated by the following equation (fully described in \citealt{saint}) :
\begin{equation}S/N =\frac{1000F_{c}}{W_{50}}\frac{w^{1/2}_{smo}}{rms}\label{AASN}\end{equation}
Where for W50 $<$ 400 km/s, $w_{smo} = W50 / (2 \times v_{res})$, where $v_{res}$ is the velocity resolution in km/s, and for W50 $>$ 400 km/s, $w_{smo} = 400 / (2 \times v_{res})$. $F_{c}$ is the total flux in Jy; $rms$ is the rms across the spectrum in mJy. \citealt{aavirgo} estimate a S/N value of 4.5, by this equation, as a lower limit for their detections.

Figure \ref{HIMFs} shows evidence for differences in the estimates of the HIMF by each survey. At the high mass end ($>$ 9 log($M_{\rm \odot}$)), AGES detects relatively few objects compared to ALFALFA, and none at all in the highest-mass bin. Since there is no upper limit on the mass sensitivity of AGES, this discrepancy must be due to variation within the cluster. As predicted by \citealt{schneider}, large-area, shallow surveys are better for detecting high mass galaxies.

Conversely the two surveys disagree in the opposite sense at the low-mass end. While their estimates are consistent within the 7.5 bin, at 7.0 ALFALFA detects no objects, whereas AGES still detects a significant number of galaxies. Given the much larger area of coverage of ALFALFA, this result cannot be attributed to cosmic variance, but is a result of the greater sensitivity of AGES. If ALFALFA were to detect the same fraction of low-mass galaxies as AGES, it should have recorded about 20 objects in the lowest mass bin.
 
\subsection{Morphological evolution and the cluster environment}
\label{VC2Dist}

In Paper V we presented an analysis of the spatial and velocity distribution of our detections together with their colours, $M_{\rm HI}$/$L_{\rm g}$ ratios and HI deficiency estimates. The sample in VC2 is too small to repeat this analysis, but there are nonetheless some broad trends which bear  discussion.

\begin{figure}
\begin{center}
\includegraphics[width=84mm]{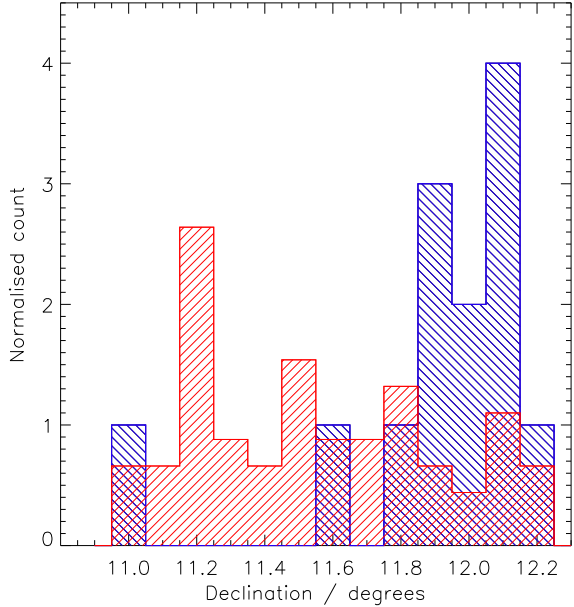}
\caption[Declination distribution of galaxies in VC2]{Declination distribution of VC2 galaxies. Red (lines slanted forwards) indicates VCC galaxies undetected in HI while blue (lines slanted backwards) indicates AGES detections. The bin size is 0.1 degrees. The mean declination of the HI detections is +11.89$\pm$0.08$^{\circ}$ (standard error); for the non-detections it is +11.55$\pm$0.04$^{\circ}$.}
\label{DecHist}
\end{center}
\end{figure}

\begin{figure}
\begin{center}
\includegraphics[width=84mm]{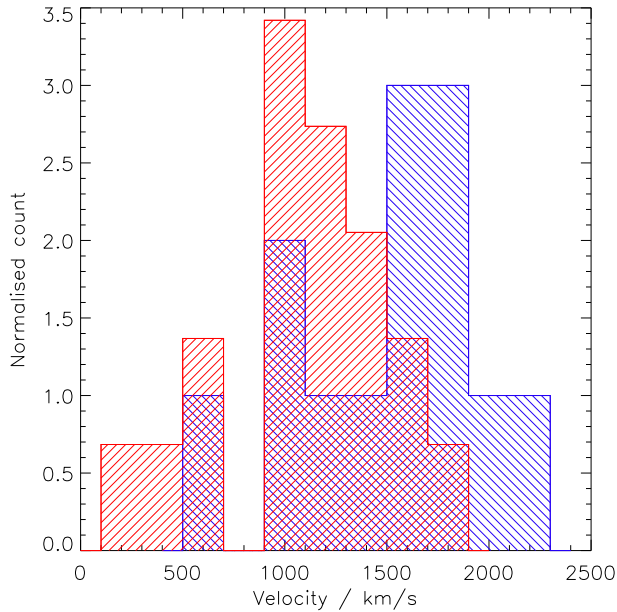}
\caption[Velocity distribution of galaxies in VC2]{Velocity distribution of VC2 galaxies, with the colour scheme as for figure \ref{DecHist}. The bin size is 200 km/s. The mean velocities are 1,513$\pm$125 km/s and 1,136$\pm$50 km/s for the detections and non-detections respectively.}
\label{VelHist}
\end{center}
\end{figure}

As there is supposedly only a population of galaxies at a single distance in the VC2 region, we might not expect to see much difference in the distribution of HI detected and non-detected galaxies. In fact the distributions are quite different, both in terms of spatial positions (unlike in VC1), as shown in figure \ref{DecHist}, and velocity (figure \ref{VelHist}). The HI detections are concentrated in a narrow band in declination, whereas the non-detected VCC galaxies are distributed more uniformly in declination. The HI detections are found preferentially at higher velocities than the non-detections, a similar result to that found in Paper V. These results suggests that this part of the cluster is not comprised only of galaxies at a single distance as was previously thought. This is further evidence that the individual subclusters of Virgo are themselves still assembling (see paper V). 

\begin{figure*}
\begin{center}
\includegraphics[width=168mm]{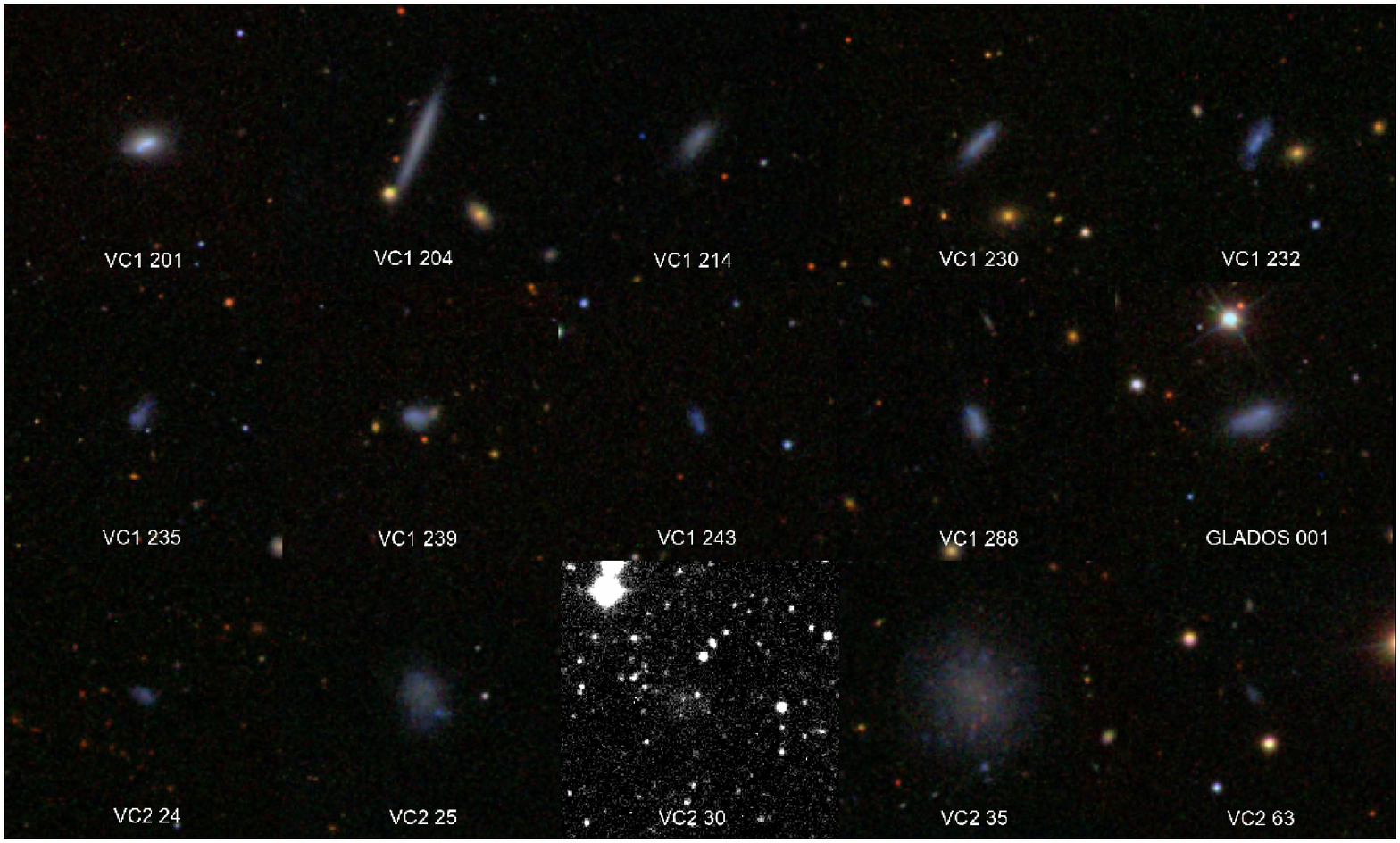}
\caption[SDSS images of galaxies detected by AGES not listed in the VCC, for both regions]{Optical images of galaxies detected by AGES not listed in the VCC. Images are all 1.7$\arcmin$ across. For the exact selection see the text. For VC2 30 a $B$-band image from the INT WFS is used, for all others SDSS RGB images are shown. AGESVC2\_25 and AGESVC2\_63 are outside the VCC area.}
\label{NonVCCs}
\end{center}
\end{figure*}

Five galaxies in VC2 are detected which are not members of the VCC, though two of them are outside the VCC area. As with many of the non-VCC detections in VC1, most are small, optically faint dwarf irregulars best described as fuzzy blue blobs in optical images. We show finding charts for a sample of these detections in figure \ref{NonVCCs}. For VC1 we show only those objects with no previous HI detection and a single sure or probable optical counterpart. For VC2 we show all five of our non-VCC detections.

Many of the non-VCC detections resemble edge-on discs, an unexpected result since these should have higher surface brightness than face-on discs (and thus be easier to detect in an optical survey). It is possible that as they are relatively faint, they were mistaken for background spirals.  All the detections are late-type. As we showed in VC1, these objects are likely to be infalling from the field since they are small and retain their gas content. This is reinforced by the presence of giant spirals with strong HI deficiency. The ability of small, gas-rich objects to survive in an environment where much more massive objects have lost nearly all of their gas strongly suggests that they are relatively recent arrivals to the cluster.

In the VC1 region it was found that most HI detections were of late-type objects, even if not members of the VCC, with virtually all late-type VCC objects being detected. This is also the case in VC2. There are only two undetected unclassified objects - VCC 1977 and 2085. Both are so faint ($m_{g}$ $<$ 18.0) and red ($g - i$ $>$ 1.0) that they are probably dwarf ellipticals. Thus every late-type VCC cluster member in this region is detected in HI. 

While every late-type galaxy is detected in HI, not every HI detection corresponds to a late-type galaxy. In VC1 about 7$\%$ of the HI sample correspond to early-type objects, and it was suggested that the detected S0s are being evolved via gas loss from spirals and the dEs are recently accreted field objects. In VC2 there are only 13 HI detections so if the detection rate of early-types is consistent with that of VC1, a single early-type HI detection might be expected. 

In fact there are two possible detections of ETGs detected in HI, a dwarf elliptical and an S0 linked by an HI stream. The VCC 2062/2066 system has previously been described in \citealt{vv89} and \citealt{duc07}, who used the Westerbork and VLA interferometers to achieve much higher resolution than is possible at Arecibo. While AGES does cover a larger area, we do not see any more extended emission around the system.

We also confirm that there is far more gas in this system than would be expected from a dwarf irregular : \citealt{vv89} estimate that VCC 2062 would have had a $M_{\rm HI}$/$L$ ratio of 40 if it was the sole source of the gas. However if all of the gas was originally associated with VCC 2066, which is much brighter, the $M_{\rm HI}$/$L$ need only have been 0.03, consistent with the other optically bright S0 objects detected in HI by AGES in the cluster. VCC 2066 could be a further possible example of an S0 evolving (in part via gas loss) from a spiral galaxy.

As discussed in Paper V the $M_{\rm HI}$/$L_{\rm g}$ ratio is closely linked to HI deficiency - how much gas a galaxy has lost compared to a field object of the same morphology and optical diameter. Deficiency is calculated from the relation :

\begin{equation}HI_{def} = log(M_{HI,ref}) - log(M_{HI,obs})\label{DefEqt}\end{equation}
Where HI$_{def}$ is the HI deficiency, M$_{HI,ref}$ is the HI mass of the reference galaxy, and M$_{HI,obs}$ is the HI mass of the observed galaxy. The expected HI mass can be calculated by a linear equation, using the parameters of \citealt{b09} :
\begin{equation}log(M_{HI,ref}) = a + b\times log(d)\label{HIref}\end{equation}
Where $a$ and $b$ depend upon the morphological type and $d$ is the optical diameter in kpc. A spiral galaxy with the optical diameter of VCC 2066 would require a deficiency of 1.1 to have the same gas content. Thus, despite being optically bright and relatively HI massive, VCC 2066 is not inconsistent with the S0s detected in HI in Paper V.

\citealt{bg06} conclude that the formation of S0s via ram-pressure stripping of spiral galaxies is unlikely. Their principle argument is that ram-pressure should not be able to change the bulge-to-disc ratio of a galaxy, yet S0s have systematically larger such ratios than spirals. They note that this change is more consistent with a tidal interaction. Yet there is no reason to suppose that the two effects do not coexist - a galaxy which experiences gas loss (by any mechanism) will still be subject to the same tidal interactions as a galaxy which does not.

There are no detections in VC2 without obvious optical counterparts, unlike in VC1 which we discuss in detail below. However AGESVC2\_30 has an exceptionally high $M_{\rm HI}$/$L_{\rm g}$ ratio of close to 10, with an estimated deficiency of 0.80. If this estimate is correct, its original $M_{\rm HI}$/$L_{\rm B}$ ratio (i.e. before it entered the cluster, assuming it to have the expected HI mass that the deficiency calculation implies) could have been much higher, up to 80 if there was no optical evolution. Thus, given its S/N of more than 16, if this object really is deficient and no optical evolution has occurred, it implies the existence of a population of infalling, non-deficient objects with very high $M_{\rm HI}$/$L$ ratios. A major caveat is that that deficiency is not well calibrated for dwarf galaxies.

\subsection{The Tully-Fisher Relation}
\label{sec:TF}
\begin{figure}
\begin{center}
\includegraphics[width=84mm]{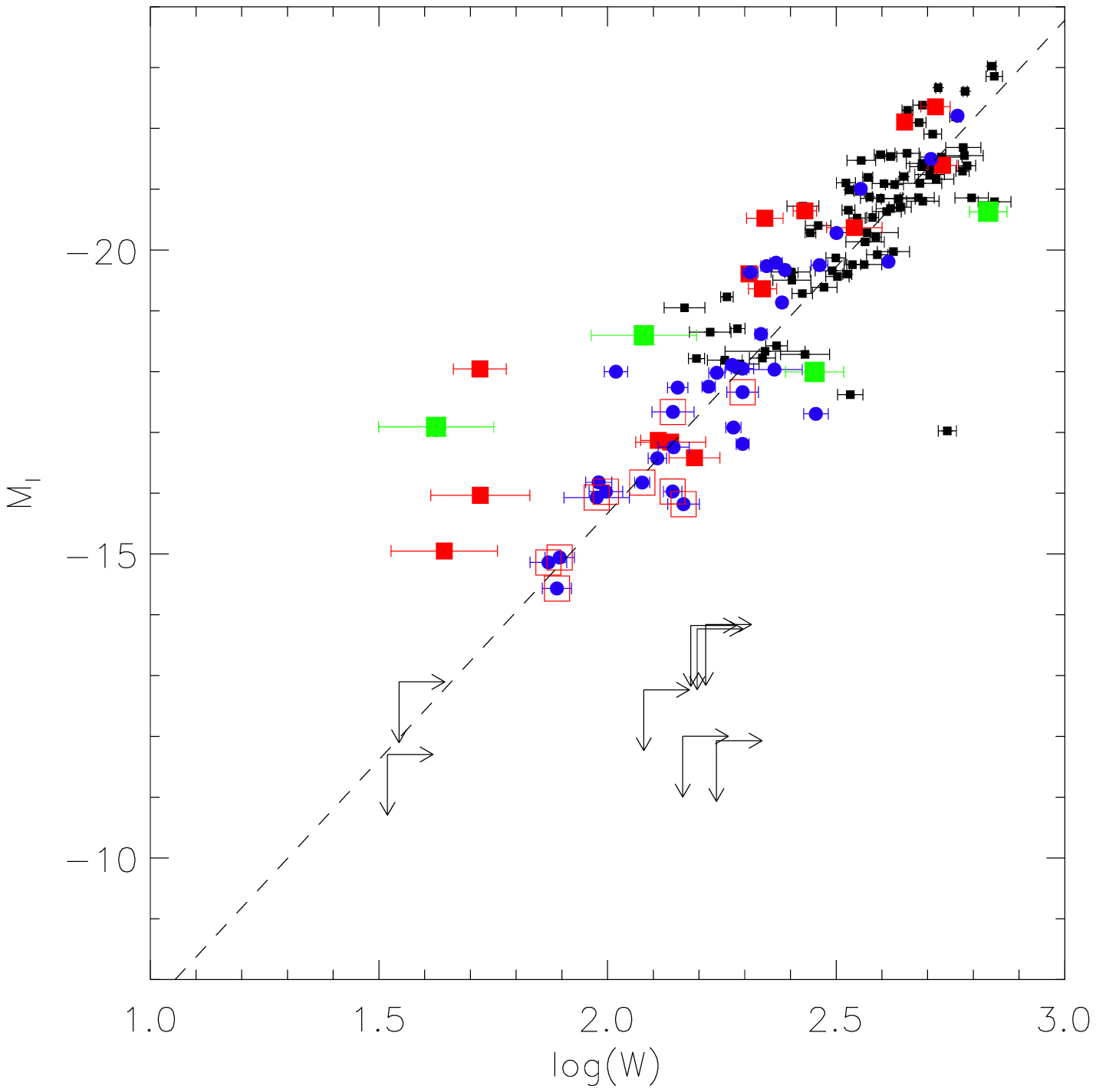}
\caption[TF]{Tully-Fisher relation using the $W_{20}$ velocity estimator. Black squares are detections behind the cluster. Blue circles are detections within the cluster with deficiency $<$ 1.0, red squares are cluster detections with deficiency $>$ 1.0. Green squares are early-type galaxies, while open red squares highlight cluster detections which are not listed in the VCC. Upper limits are shown for objects without obvious optical counterparts (see text). Here the dashed line is the TF relation determined in \citealt{Tully}.}
\label{TF20}
\end{center}
\end{figure}

\begin{figure}
\begin{center}
\includegraphics[width=84mm]{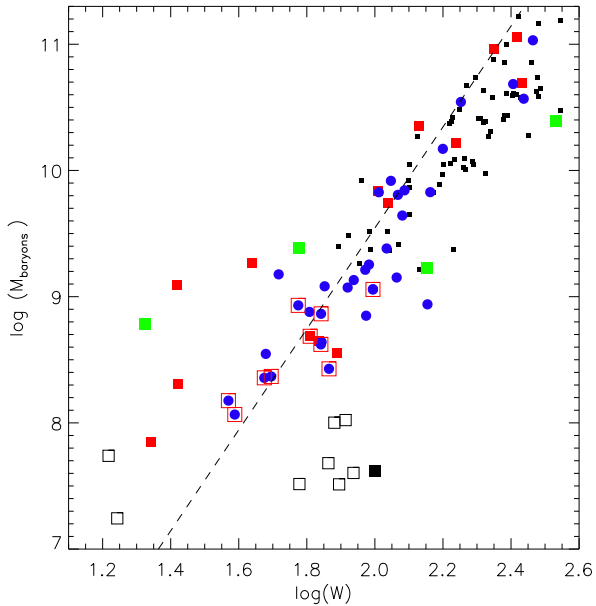}
\caption[TF]{Baryonic Tully-Fisher relation where M$_{baryon}$ = 1.4$\times$M$_{HI}$ + M$_{*}$, following \citealt{MG00}. The dashed line is the best-fit determination from \citealt{MG00}. The colour scheme is as for figure \ref{TF20}, except that our optically unseen detections are shown as open black squares. We also include VIRGOHI21 as measured in \citealt{m07} (filled black square).}
\label{BaryonicTF}
\end{center}
\end{figure}

\begin{figure}
\begin{center}
\includegraphics[width=84mm]{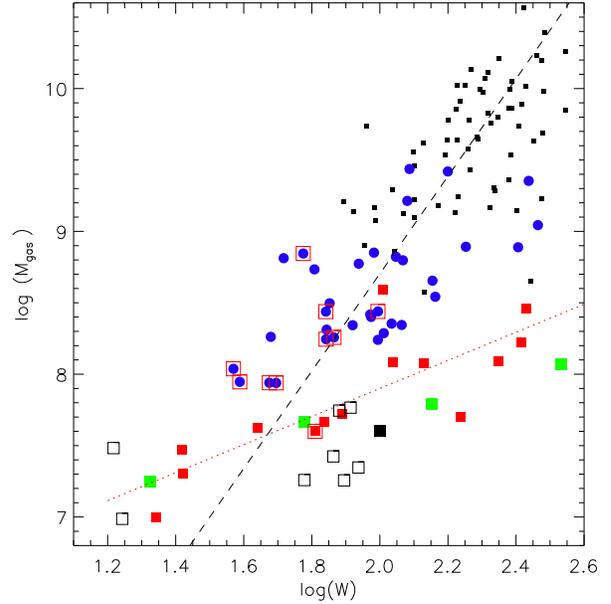}
\caption[TF]{Gas-only Tully-Fisher relation where M$_{gas}$ = 1.33$\times$M$_{HI}$, following \citealt{MG12}. The dashed line is the best-fit determination from \citealt{MG12}. The red dotted line is the best-fit to the strongly HI-deficient galaxies. The colour scheme is as for figure \ref{BaryonicTF}.}
\label{GasTF}
\end{center}
\end{figure}

While the Tully-Fisher relation is traditionally used as a distance estimator, it also offers insights into galaxy evolution. In figure \ref{TF20} we plot the Tully-Fisher using conventional $I$-band luminosity (using the simple correction $I = i - 0.75$, \citealt{win}). In figure \ref{BaryonicTF} we plot total baryonic mass rather than luminosity, following \citealt{MG00}. In figure \ref{GasTF} we use only the gas mass, following \citealt{MG12}. In all plots we have corrected our rotation widths for inclination angle (for the baryonic and gas-only relations we assume that $V_{circ} = W_{20}/2.0$, in order to compare accurately with previous studies). We make the simplistic assumption of a thin circular disc, using only objects where $i$ $>$ 30$^{\circ}$, and do not correct for extinction effects.

\begin{figure}
\begin{center}
\includegraphics[width=84mm]{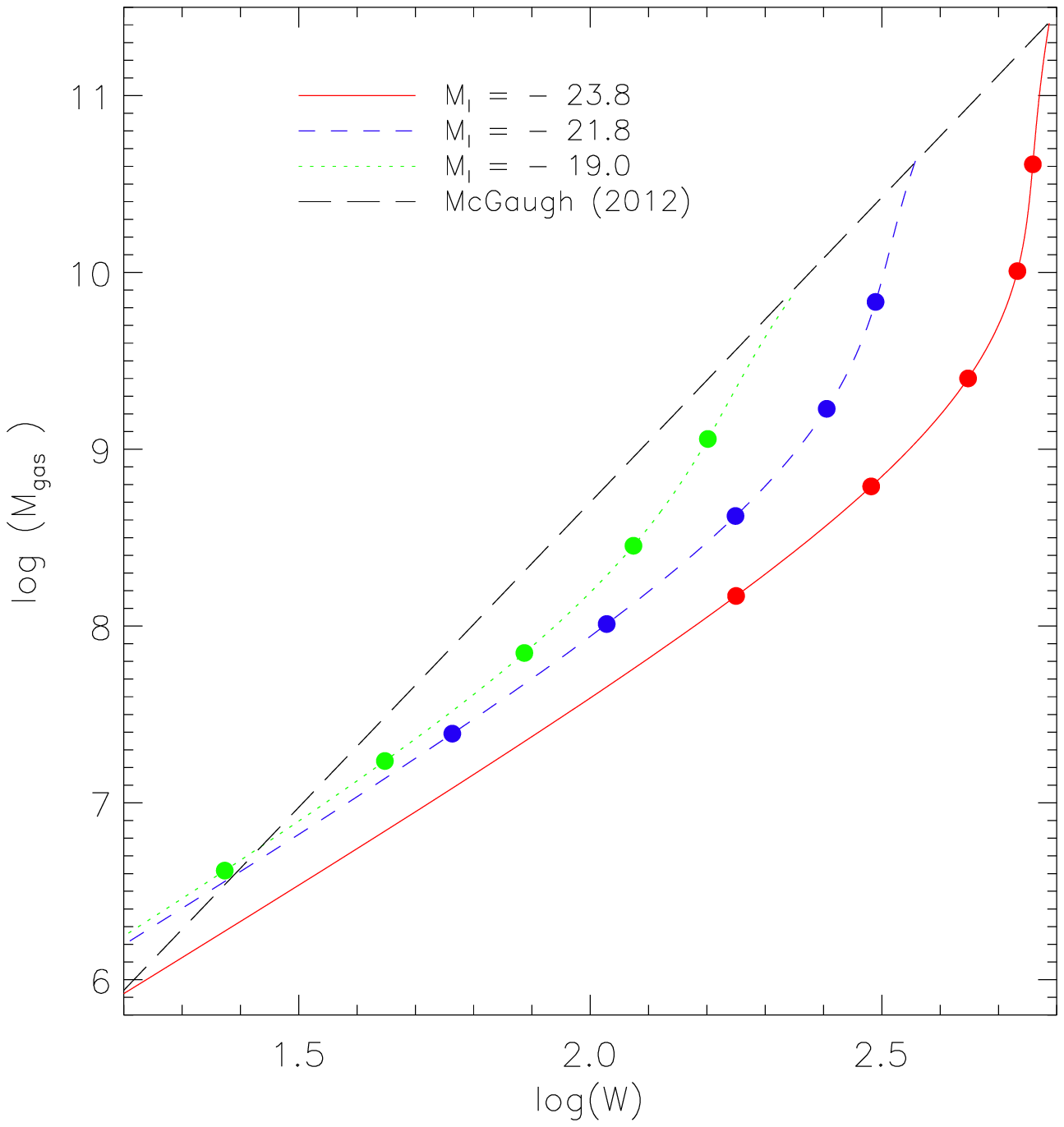}
\caption[TF]{Predicted deviation of galaxies (of three different luminosities) from the gas-only Tully-Fisher relation. Their initial gas content is that predicted by the TFR. Gas loss is complete beyond an increasing truncation radius, generating the tracks visible. Moving down the tracks, the dots indicate where the truncation radius is 40\%, 20\%, 10\%, 5\% and 2.5\% of the original radius.}
\label{Model}
\end{center}
\end{figure}

In figure \ref{TF20} we see that the background detections and the non-deficient cluster detections agree well with previous measurements of the relation from field samples. The highly deficient objects do not fit the relation, especially the less massive objects. It appears that in these latter cases the outer, most rapidly rotating gas has been removed, so that the gas is no longer tracing the true rotation curve. In effect the measurable rotation curve of the gas has been truncated (\citealt{gv08} note that this is consistent with ram-pressure stripping). The brighter deficient galaxies appear to be able to maintain their line width while still losing significant amounts of gas, which may relate to the different shape and extent of the rotation curve for objects of different masses (\citealt{cat}). More massive galaxies have a greater fraction of gas in the flat part of their rotation curve, whereas low-mass galaxies can have curves which are always rising.

An alternative interpretation is that galaxies deviant from the TFR are actually at a different distance and are seen in projection against the Virgo cluster. For the most deviant objects, this would require them to be at a distance of 2 Mpc in order to reconcile their measured and TFR-predicted absolute magnitudes. This is completely at odds with their velocities (1,000-2,000 km/s), so a truncated gas disc, and correspondingly reduced line width, seems a far more likely explanation (we discuss this further below).

For those objects in VC1 which we deemed to have no obvious optical counterpart, we chose the brightest candidate smudges visible in the optical images. We use the HI velocity widths as measured without attempting to correct for inclination effects (since the only optical candidates visible are extremely faint). The positions shown in figure \ref{TF20} are thus lower limits in terms of velocity widths. Since we used the brightest visible objects, they are also lower limits in terms of optical magnitude. Note that although some of the non-VCC HI detections are also extremely optically faint (figure \ref{NonVCCs}), they all lie on the Tully-Fisher relation : six of the objects which we have identified as having no optical counterparts are categorically different.

This is also supported by the baryonic TFR in figure \ref{BaryonicTF}. Here we also plot VIRGOHI21, an object variously claimed to be a possible primordial `dark galaxy' (\citealt{m07}, \citealt{d04}), a product of harassment (\citealt{AAvhi}), and the result of a single high-velocity interaction (\citealt{duc08}). Both VIRGOHI21 and 6 of our detections without obvious optical counterparts (again, we use the brightest optical candidates which are visible) have nowhere near enough baryons - by a about a factor of 10 - to lie on the TFR. Again this is quite different to the cases where the counterparts are merely unusually faint, even though those objects have high M$_{HI}$/L ratios. Additionally, we note that galaxies with low surface brightnesses have been shown to agree with the TFR of high surface brightness galaxies (\citealt{Zwaan}). By extension, objects which have \textit{no} optical surface brightness should also be expected to obey the baryonic TFR.

The natural inference is that the optically undetected objects cannot be primordial gas clouds embedded in dark matter halos - otherwise we would expect them to follow the same TFR as all the other faint objects do. There is a caveat in that this assumes objects which never form any stars at all are fundamentally dynamically similar to objects which do. A counter-argument might be that objects which cannot form stars must have lower gas densities, so we might expect them to lie below the normal TFR by definition (for example see \citealt{d06}).

Since none of the optically undetected objects have spatially resolved HI components, it is difficult to estimate their dynamical masses (as this requires knowledge of how far the HI traces the rotation curve). However, we can offer a crude estimate. \citealt{leroy} find that dwarf galaxies exhibit a very narrow range of gas surface densities, from 5-10 $M_{\rm \odot}$ pc$^{-2}$. Assuming a typical value of 7.5 $M_{\rm \odot}$ pc$^{-2}$, the six detections which lie below the TF would have gas disc radii ranging from 0.8-1.6 kpc (following \citealt{leroy} this accounts for helium by  $M_{\rm gas}$  = 1.36$\times$$M_{\rm HI}$). This in turn leads to dynamical to gas mass ratios ranging from 20-60, assuming them to be virialised. Thus they are either highly unstable, transient clouds of gas, or stable but optically inert systems. We cannot offer further insights without resolved HI mapping (which would not only determine the true gas disc radii, but also show if there was ordered rotation).

We note that two of the detections without optical counterparts are not found in the same parameter space in the TF diagram as the others. Rather than falling below the TFR, the two narrowest detections lie above it. In one sense, it is not even necessary to assign them optical counterparts - their HI mass alone is sufficient for them to slightly exceed the baryonic mass predicted by the TFR. Since we cannot estimate their inclination angle it is possible that their true velocity widths are much higher, but we cannot rule out the possibility that these really are primordial, optically inert objects.

Making the same assumptions as for the other objects, these two objects require dynamical to gas mass ratios of 2-3, far less than for those objects which do not satisfy the baryonic TFR. In essence they are conceivably massive enough to be gravitationally bound without a dark matter component. This adds some credence to the possibility that they are tidal debris - if they are self-bound, they may survive long after any more extended, unbound tidal features have dissipated.

Figure \ref{GasTF} plots the TFR using only the gas mass. While this has significantly more scatter, several other features become apparent from this plot. The most deficient galaxies are seen to follow a different slope to the non-deficient objects, and both the ETGs and optically unseen HI detections broadly follow this trend. The highly deficient objects with extremely low and extremely high rotation widths imply that gas removal can have different effects on galaxies, depending on their total mass.

Firstly, the deficient objects which have extremely low velocity widths ($\log(W)$ $<$ 1.6) have predicted HI masses (see equation \ref{HIref}) as high as 9.0 in logarithmic solar units. If these objects have not experienced any reduction in their line width, this implies they were originally well above the TFR. Yet none of the non-deficient cluster objects - or background objects - lie in this region of the TF diagram. This implies that these objects must originally have had much broader line widths in order to obey the predictions from the TFR.

Secondly, the opposite is true for the deficient objects with high velocity widths ($\log(W)$ $>$ 2.2).  Their predicted original HI masses (from equation \ref{HIref}) are as high as 9.8 (logarithmic solar units), which would bring them into close agreement with their TFR predictions - but only if their velocity widths were not significantly altered. If they originally had much broader velocity widths, as the low-mass deficient objects seem to have done, then they would lie well below the TFR - a region where no non-deficient objects are actually observed. A caveat is that sensitivity would prevent the detection of even very HI-massive galaxies with sufficiently broad lines, but this would require absurdly high velocity widths. At 100 Mpc, a galaxy with an HI mass of 10.0 would require a line width of 2,300 km/s to reach a S/N of 3.0. Therefore, very massive galaxies which lose even 90\% of their original gas content probably do so without significant truncation of their measurable rotation curves.

We create an analytical model to verify if ram pressure stripping can account for the observed deviation of deficient galaxies from the TFR. We assume rotation curves as given in \citealt{cat} for galaxies of three different luminosities. Each model galaxy has a gas disc that is initially large (extending out to 6 exponential disc radii). As the model's disc is ram pressure stripped the gas mass falls, and its disc is truncated to a truncation radius $r_{\rm{trunc}})$, such that fast moving gas in the outer rotation curve will no longer feature in the observed line width. The line width is assumed to be twice the circular velocity and the gas disc to be of constant surface density (as is typically observed - for example \citealt{w08}).

The initial mass of gas in the disc is chosen to be such that each galaxy initially lies on the \cite{MG12} trend. To calculate how the gas mass changes, the truncation radius of each model's HI disc is reduced in small steps. The gas mass and line width is recalculated at each step, thus creating the galaxy tracks in figure \ref{Model}.

The most luminous galaxy track initially fall vertically on the plot as its gas is stripped, but its line width changes little. This is due to the shape of the rotation curve - while some of the `flat' part of the rotation curve remains, the line width does not fall significantly. However once the flat part has been stripped, the line-width falls rapidly, and the track veers to the left. The least luminous galaxy model does not have the initial vertical drop at all, as its rotation curve does not really have a flat outer part, but instead continuously rises slowly.  Quantitatively the agreement is not ideal between observed deficient galaxies and the model - perhaps as a result of our rather idealised assumptions. However qualitatively, the \cite{MG12} line and HI deficient galaxies cross at low $\log(W)$ while the HI deficient galaxies are also found increasingly far below the \cite{MG12} line with increasing $\log(W)$. Therefore a ram pressure scenario can indeed reproduce the qualitative behaviour.

\section{Summary and Discussion}
\label{sec:Conclusions}

VC2 is a poorer region than VC1 both in HI and optical, with a factor of 1.8 fewer HI detections per square degree in VC2 compared to VC1. Other regions within the cluster show even greater differences, so the difference is low enough to be readily explained by normal variance within the cluster. The number of background HI detections in each region is consistent. The low number of detections in this region is therefore not a result of poor source extraction methods, despite the high S/N level of the detections. Indeed, the HI mass distribution from AGES shows a greater fraction of objects in the low-mass bins than other shallower surveys.

Unlike in VC1, the spatial and velocity distributions of the AGES HI and VCC optically-detected galaxies are clearly different. The HI detections are clustered in declination (whereas the VCC non-detections are uniformly distributed) and tend to be found at higher velocities within the cluster. This suggests that subcluster A is still assembling, with the HI detections being recently accreted from the field - perhaps from a single infalling group, since their spatial and velocity distributions are quite different to and distinct from the HI non-detections.

In Paper V we described evidence for a variation in HI deficiency with radial distance from the cluster center. The low numbers prevent a robust comparison in VC2, though the two most deficient galaxies are found in the western part of the cube, closer to the cluster centre. Galaxies of lower deficiency are found in the same region, as is the case in VC1. This supports the view that the cluster is still assembling, with the low-deficiency objects closer to the cluster centre experiencing their first pass through that region. It is difficult to see how such faint objects as VC2\_30 could otherwise have survived in an environment where  bright objects such as AGESVC2\_033 (VCC 1859), which has an M$_{g}$ of -18.7, have lost almost all of their gas (deficiency of 1.8).

In Paper V the detection of several S0 galaxies with low masses of HI was interpreted as evidence for morphological evolution from spirals to lenticulars (i.e. from one Hubble type to another). In VC2 the statistics are too poor to comment further, except to note the detections of VCC 2066. As discussed above, this is an extremely unusual system which, for a proper understanding, requires a detailed study in its own right (see \citealt{duc07}). We have also noted that the most deficient galaxies tend to have lower velocity widths than predicted by the Tully-Fisher relation. Two ETGs are also offset from the Tully-Fisher, which may be further evidence of gas-loss driven morphological evolution.

Some strongly HI deficient galaxies appear to be quite normal late-types in most other respects (they do not, for instance, show signs of any unusual morphologies or extended HI features). Indeed, some detections are of much lower $M_{\rm HI}$/$L_{\rm g}$ than field objects, but of exactly the same colour. Morphological and/or colour evolution may not be driven solely by gas removal (e.g. the harassment scenario of \citealt{m99}).

The HI detections within the cluster that are not in the VCC are all generally small, blue, gas-rich objects. Those in VC2 are all non-HI deficient, except for object 30, but some in VC1 have deficiencies greater than 0.5. A caveat is that the HI deficiency relation is not well calibrated for objects other than spirals (\citealt{sol96}). More complex procedures to determine the HI deficiency may eventually improve the situation for dwarfs, but these have so far only been attempted on very small samples (\citealt{lee}).

There are no dark galaxy candidates anywhere in VC2, including the background volume. Several low S/N candidates were originally detected but all were found to be spurious on follow-up. The closest object to a dark galaxy is object 30, which has the highest $M_{\rm HI}$/$L_{\rm g}$ ratio in the sample at almost 10, but there is nonetheless a obvious possible optical counterpart at the coordinates of the HI. Identifying objects with very faint or non-existent optical counterparts is difficult and subjective, but use of the Tully-Fisher relation may help. Six detections in VC1 are found for which the only possible optical counterparts would lie well to the right of the Tully-Fisher : they are much fainter than expected from their velocity widths. 

These detections correspond to objects we labeled as being optically dark in paper V, suggesting that the optical counterparts which are visible are not really associated with the HI. Furthermore, combining the detected stellar and gas emission is not sufficient to bring these objects into agreement with the prediction from the baryonic TFR, based on their velocity widths. Nor do they have even sufficient gas to fit a gas-only TFR. If they are in virial equilibrium, they require as much as 60 times the amount of dark matter as gas in order to be gravitationally bound.

However, whatever these objects actually are, it seems very unlikely that they are primordial, optically dark galaxies. Those objects which have readily identifiable optical counterparts agree with the TFR predictions no matter how optically faint they are. Highly deficient objects are an exception to this, but they deviate from both the stellar and baryonic TFR in the opposite sense to the optically undetected objects. Deficient objects have unexpectedly low velocity widths given their mass (both stellar and baryonic), whereas optically undetected objects have unexpectedly high velocity widths given their mass, with respect to the TFR.

The most obvious alternative is that the objects are remnant gas removed from ordinary galaxies. None of the detections here show any obvious signs of extended HI features which would allow us to identify possible initial sources of the gas clouds. It is, of course, possible that any such streams are below our sensitivity limit (in the manner of the stream which includes VIRGOHI21 - \citealt{AAvhi}). While five of our detections are in close proximity ($<$ 30\arcmin) to other objects, this is not in itself necessarily significant, given the galaxy density of the cluster. Only deeper, higher spatial resolution observations will be able to offer further insight on this issue.

As in VC1, no new extended HI features are detected in the cluster. The known HI complex associated with VCC 2066 and 2062 is detected, as are several HVCs. This means the lack of extended HI detections is not due to data reduction or source extraction procedures. Although it is clear that some Virgo galaxies are significantly HI deficient, it remains an open question as to where this HI actually goes. Very extended HI features, such as VIRGOHI21 (\citealt{m07}, \citealt{AAvhi}), do not appear in any way common around gas deficient galaxies. 

While there are key differences between the VC1 and VC2 regions - different infalling clouds and different galaxy densities - in many ways they are also very similar. Both show a mix of highly deficient and non-deficient galaxies in approximately the same proportion. Both show some evidence that deficiency varies radially within the cluster. Both indicate that dark galaxies and extended HI features are, at the least, rarities within the cluster environment. Perhaps most importantly, both demonstrate that the effects of gas loss can be far from simple - highly deficient objects may share few other - if any - common features. The key result from this sample is that the effects of gas loss are not always clear, and can be difficult to disentangle from other environmental effects and the varying properties intrinsic to individual objects.

\section*{Acknowledgments}

R.T. wishes to thank Stacy McGaugh for his helpful suggestions regarding the Tully-Fisher relation and dark galaxies.

This work is based on observations collected at Arecibo Observatory. The Arecibo Observatory is operated by SRI International under a cooperative agreement with the National Science Foundation (AST-1100968), and in alliance with Ana G. Méndez-Universidad Metropolitana, and the Universities Space Research Association. 

This research has made use of the GOLDMine Database. 

This research has made use of the NASA/IPAC Extragalactic Database (NED) which is operated by the Jet Propulsion Laboratory, California Institute of Technology, under contract with the National Aeronautics and Space Administration. 

This work has made use of the SDSS. Funding for the SDSS and SDSS-II has been provided by the Alfred P. Sloan Foundation, the Participating Institutions, the National Science Foundation, the U.S. Department of Energy, the National Aeronautics and Space Administration, the Japanese Monbukagakusho, the Max Planck Society, and the Higher Education Funding Council for England. The SDSS Web Site is http://www.sdss.org/.

The SDSS is managed by the Astrophysical Research Consortium for the Participating Institutions. The Participating Institutions are the American Museum of Natural History, Astrophysical Institute Potsdam, University of Basel, University of Cambridge, Case Western Reserve University, University of Chicago, Drexel University, Fermilab, the Institute for Advanced Study, the Japan Participation Group, Johns Hopkins University, the Joint Institute for Nuclear Astrophysics, the Kavli Institute for Particle Astrophysics and Cosmology, the Korean Scientist Group, the Chinese Academy of Sciences (LAMOST), Los Alamos National Laboratory, the Max-Planck-Institute for Astronomy (MPIA), the Max-Planck-Institute for Astrophysics (MPA), New Mexico State University, Ohio State University, University of Pittsburgh, University of Portsmouth, Princeton University, the United States Naval Observatory, and the University of Washington.

{}

\bsp

\appendix
\section{Comments on individual detections}
\begin{figure}
\begin{center}
\subfloat[Moment 0 map of VC1 cloud complex from 150 to 350 km/s. Contours are from 75 mJy/beam increasing in steps of 50 mJy/beam.]{\label{fig:edge-a}\includegraphics[width=84mm]{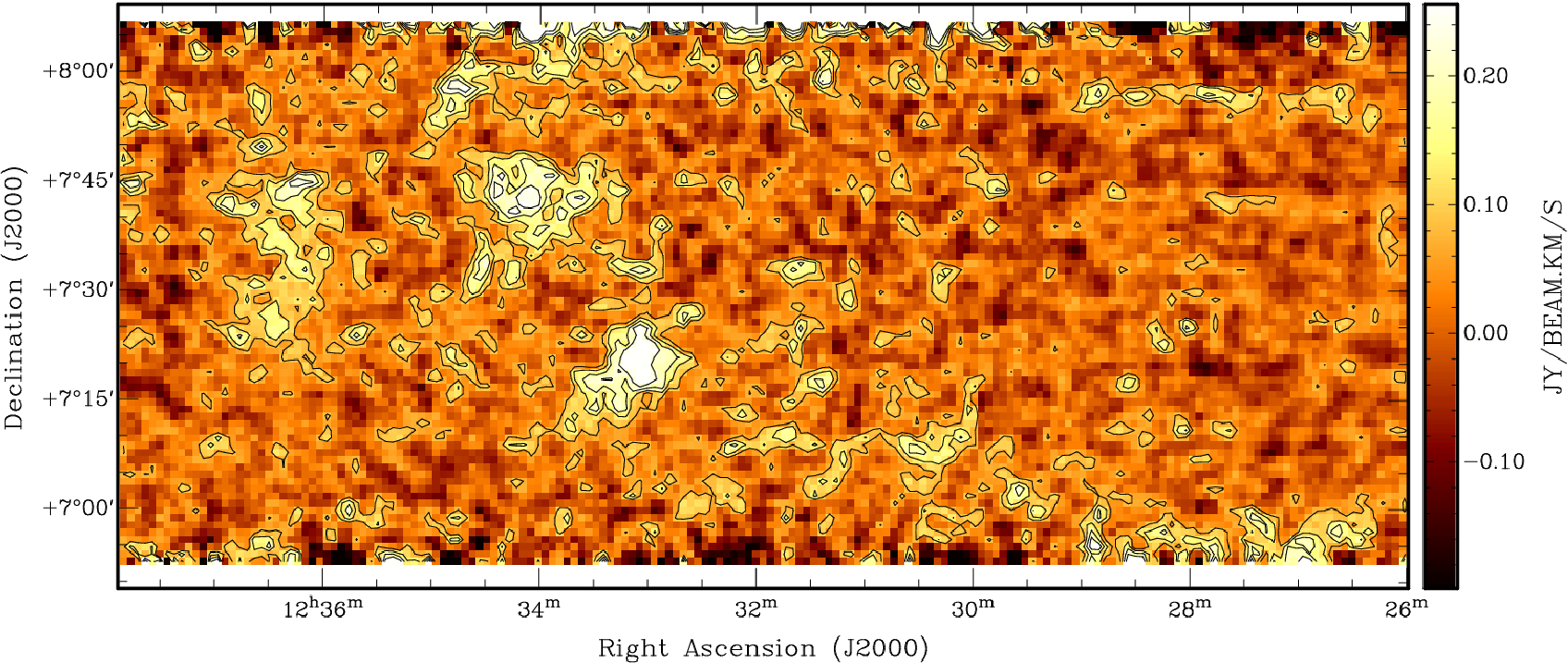}}\\
\subfloat[Moment 0 map of VC2 clouds from 220 to 280 km/s. Contours from 55 mJy/beam increasing in steps of 30 mJy/beam]{\label{fig:edge-a}\includegraphics[width=84mm]{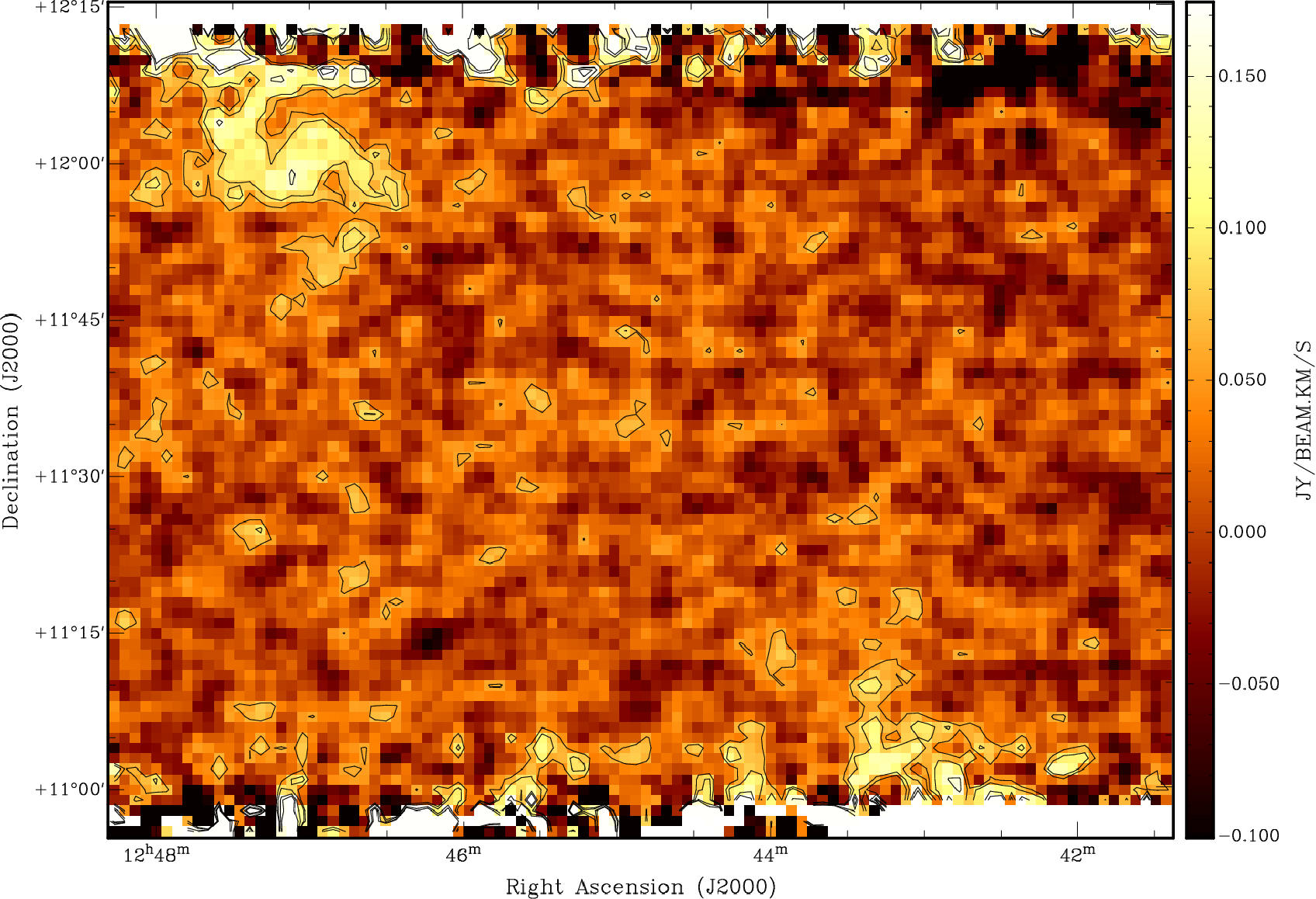}}
\end{center}
\caption[HVCs]{Integrated flux maps of probable High Velocity Clouds detected in VC1 (top) and VC2 (bottom).}
\label{HVCs}
\end{figure}

\textit{High Velocity Clouds : }A few detections in both VC1 and VC2 correspond to High Velocity Clouds (figure \ref{HVCs}), which are spatially extended but have no optical counterparts. It seems unlikely that the only significantly extended features should all lie at low velocities, if they are all cluster members. More probably these are much closer, smaller, low-mass features, which we should not expect to have associated optical emission (for example, \citealt{sblitz} searched for optical counterparts for 250 HVCs unsuccessfully).

\textit{VCC 2000 : }This object is not detected in HI, however we have discovered in the process of searching optical images of the VC2 region that it posesses a stellar tail. This is easily revealed in the deeper INT observations availble of the region, and it is also visible in the $g$, $r$ and $i$ bands of the SDSS after smoothing or rebinning, confirming it as a real structure. Though VCC 2000 has been extensively studied (243 references in NED) the tail is at approximately 27 $B$ magnitudes arcsec$^{-1}$ and it appears to have remained undetected. The tail stretches about 6$\arcmin$ from the center of VCC 2000 with a length along its curve of about 17$\arcmin$, 35 kpc at 17 Mpc distance.

\begin{figure}
\begin{center}
\subfloat[Rebinned SDSS RGB image of the region (pixels are a factor 5 larger than standard)]{\label{fig:edge-a}\includegraphics[width=40mm]{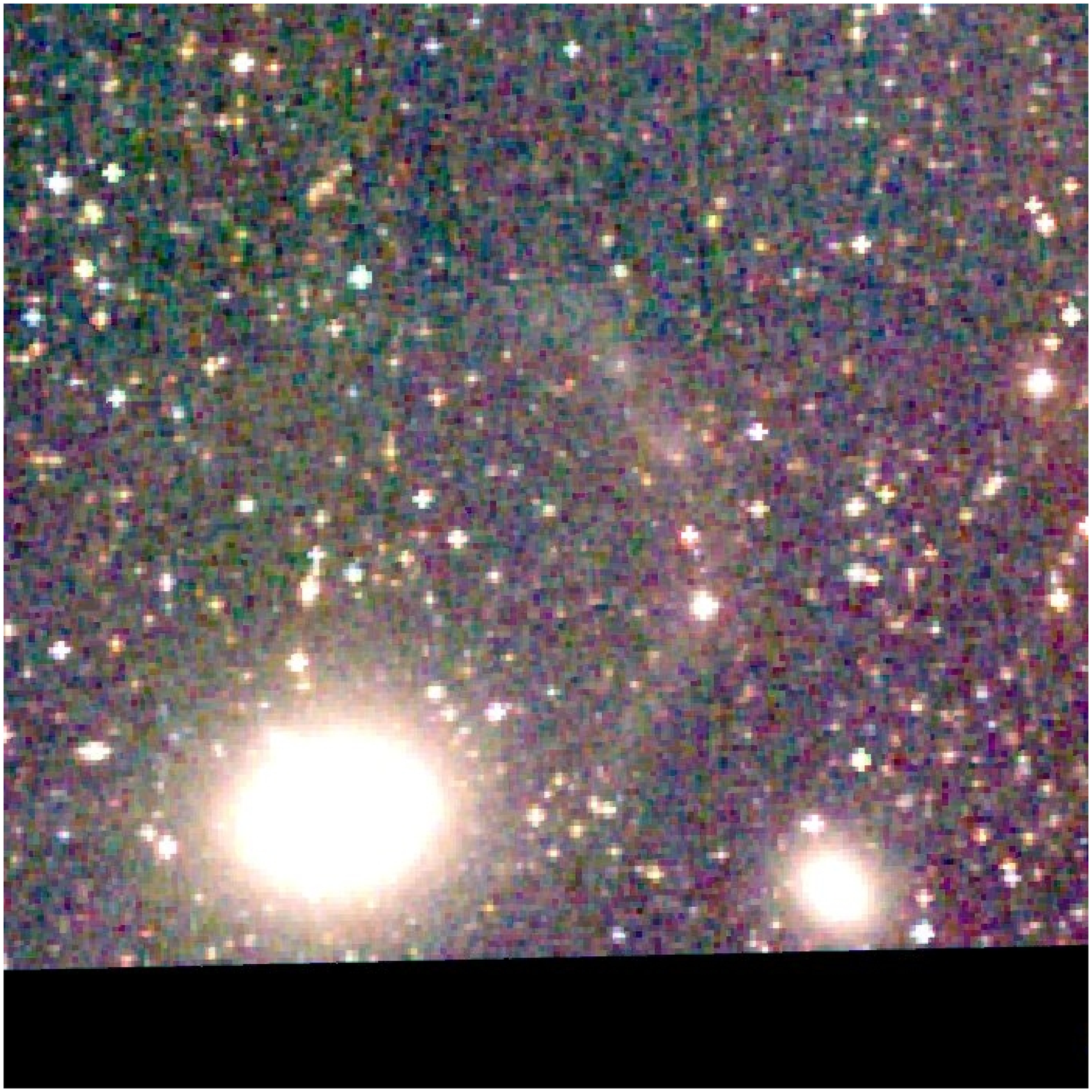}}
\subfloat[2 stacked $B$-band images from the INT WFS]{\label{fig:edge-a}\includegraphics[width=40mm]{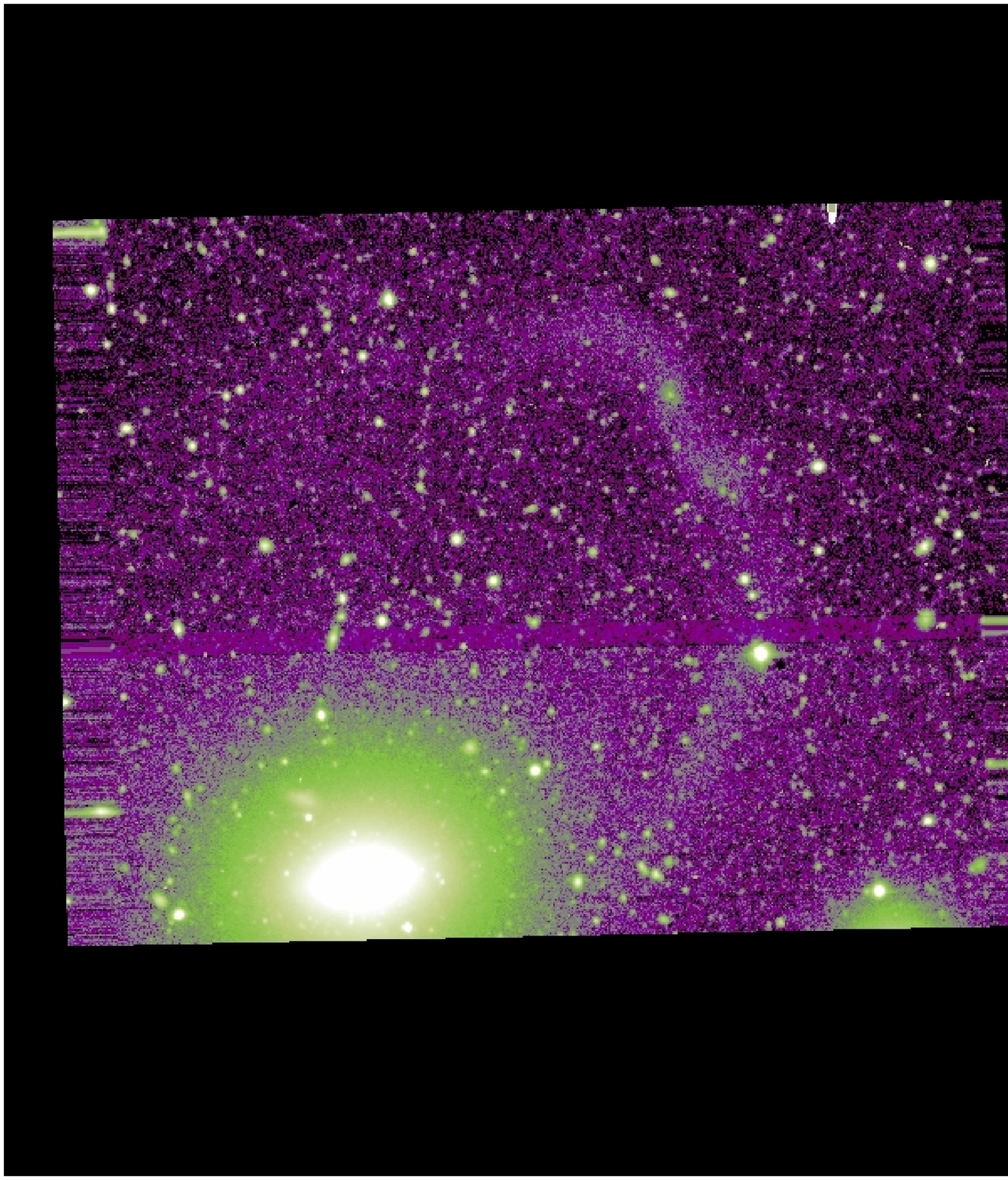}}
\end{center}
\caption[Images of VCC 2000 and environs]{Images of VCC 2095, VCC 2000 and environs, all 12$\arcmin$ across.}
\label{VCC2000}
\end{figure}

It is very difficult to accurately estimate the colour of the tail, since it is barely detected in either the SDSS $g$ band or the INT $i$ band due to fringing. From aperture photometry the $B$-$i$ colour of VCC 2000 is estimated to be 1.77$\pm$0.04, the nearby elliptical VCC 1991 to be 1.71$\pm$0.09 and the tail at 1.55$\pm$0.20. The very nearby dwarf galaxy VCC 2002 does not seem to be connected with the tail. This is confirmed by Hubble ACS images from \citealt{f06}, who note the detection of a `small, faint, very blue structure' 2.5" from the optical center apparently quite independent of VCC 2002. Unfortunately the ACS image has too small a field of view to include the tail, but it may be this is a relic of the merger of a (presumably) gas-poor dwarf.

\begin{figure}
\begin{center}
\includegraphics[width=84mm]{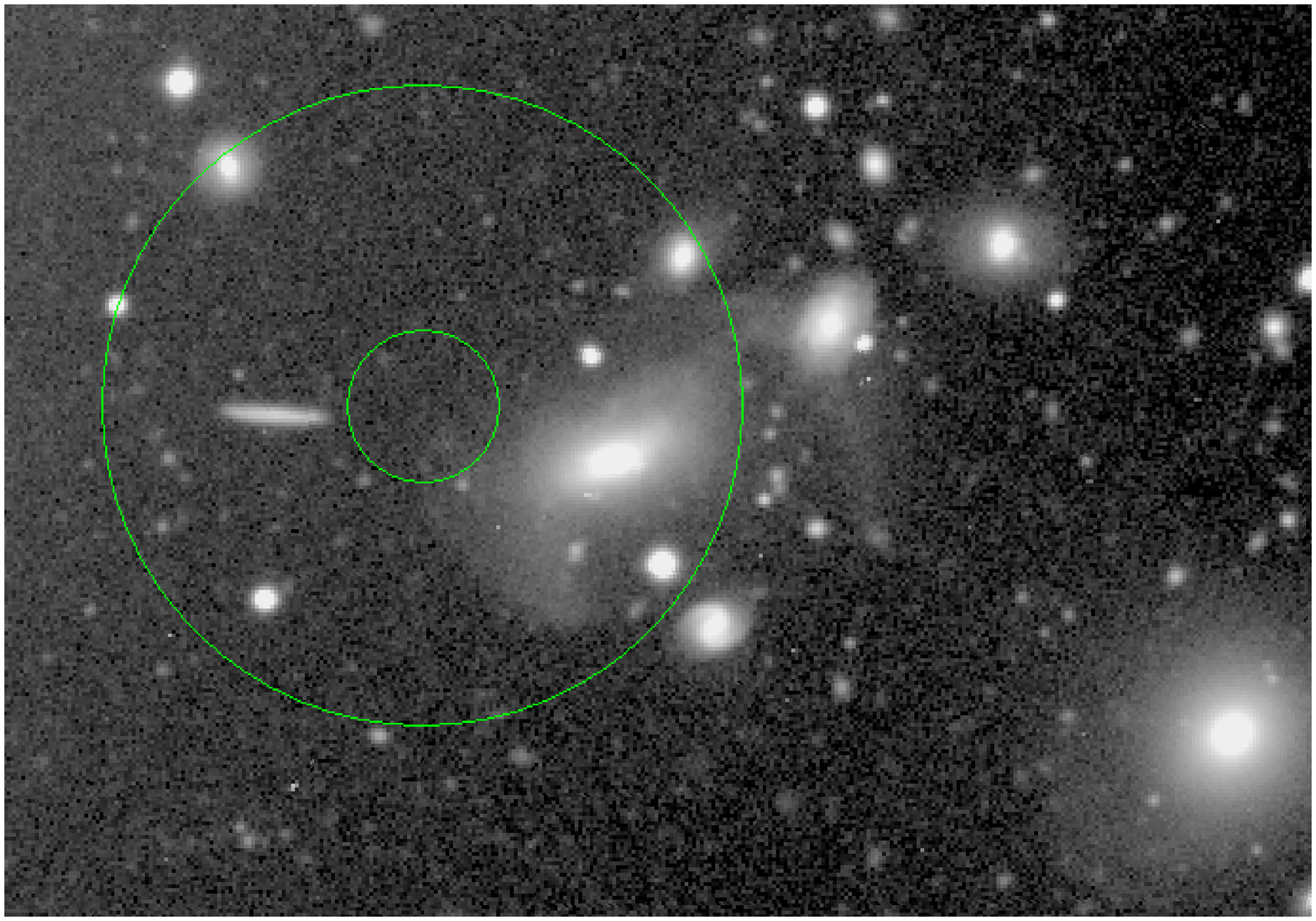}
\caption[$B$-band INT image of VCC 2041, showing tails]{$B$-band INT image of VCC 2041. The circles are centered on the coordinates of an HI detection with radii 20" and 1.4$\arcmin$; the image is about 6.6$\arcmin$ across.}
\label{2041Tails}
\end{center}
\end{figure}

\textit{VCC 2041} : In VC2 a distinct group of 5 galaxies is detected at about 13,000 km/s, all within a radius of 11$\arcmin$. The center of this group is dominated by the cD galaxy VCC 2041 and is shown in figure \ref{2041Tails} (the other 4 detections include the edge-on spiral visible to the north and 3 other late-type objects outside this image), which shows it has a disturbed optical morphology. The optical tails appear to be a new discovery - VCC 2041 has only 9 references and 1 note in NED, none of which mention its disturbed morphology. The spatial coordinates of the HI detection are midway between the elliptical and the edge-on disc (none of the other galaxies visible are HI-detected) and its redshift it also similar to both. The overall appearance of the stellar tails is perhaps suggestive of shells, which are believed to indicate recent mergers (\citealt{quin}). 

\label{lastpage}

\end{document}